\documentclass[twocolumn,showpacs,amsmath,amssymb,prd]{revtex4-1}

\usepackage{graphicx}
\usepackage{epsfig}
\usepackage{dcolumn}
\usepackage{bm}
\usepackage{mathrsfs,amsmath}

\def\beq{\begin{equation}}
\def\enq{\end{equation}}
\def\ba{\begin{eqnarray}}
\def\ea{\end{eqnarray}}
\def\<{\langle}
\def\>{\rangle}

\urldef\lensnoise\url{http://lesgourg.web.cern.ch/lesgourg/codes.html}

\begin{document}

\title{Future weak lensing constraints in a dark coupled universe} \author{Francesco De Bernardis$^1$, Matteo Martinelli$^2$, Alessandro Melchiorri$^2$,
Olga Mena$^3$ and Asantha Cooray$^1$}

\affiliation{$^1$Department of Physics \& Astronomy, University of California, Irvine, CA 92697}
 \affiliation{$^2$Physics Department and INFN, Universita' di Roma ``La Sapienza'', Ple Aldo Moro 2, 00185, Rome, Italy.}
 \affiliation{$^3$IFIC-CSIC and Universidad de Valencia, Valencia, Spain}

\begin{abstract}
  Coupled cosmologies can predict values for the cosmological parameters at low redshifts which may differ substantially from the parameters values within
non-interacting cosmologies. Therefore, low redshift probes, as the growth of structure and the dark matter distribution via galaxy and weak lensing surveys
constitute a unique tool to constrain interacting dark sector models. We focus here on weak lensing forecasts from future Euclid and LSST-like surveys combined
with the ongoing Planck cosmic microwave background experiment. We find that these future data could constrain the dimensionless coupling to be smaller than a few
$\times 10^{-2}$. The coupling parameter $\xi$ is strongly degenerate with the cold dark matter energy density $\Omega_{c}h^2$ and the Hubble constant $H_0$.
These degeneracies may cause important biases in the cosmological parameter values if in the universe there exists an interaction among the dark matter and dark
energy sectors.

\end{abstract}


\date{\today}
\maketitle

\section{Introduction}
Cosmological probes indicate that the universe we observe today possesses a flat geometry and a mass energy density made of $\sim 30\%$ baryonic plus cold dark matter
 and $70\%$ dark energy, responsible for the late-time accelerated expansion~\cite{wmap7,sdss2,snalate}. While the $\Lambda$CDM model (a flat universe with a
cosmological constant) can describe the current observational data, there exist also dynamical options for dark energy, as the quintessence fluid, in which a cosmic scalar
field is slowly approaching its ground state. This quintessence field, in principle, may couple to other fields in nature. Observations strongly constrain the
couplings to ordinary matter \cite{Carroll:1998zi}. However, interactions within the dark sectors, i.e. between dark matter and dark energy, are still allowed by
observations~\cite{Amendola:1999er,Amendola:1999dr,Amendola:1999qq,Amendola:2000uh,Amendola:2003wa,Amendola:2006dg,Valiviita:2008iv,He:2008si,Jackson:2009mz,Gavela:2009cy,
CalderaCabral:2009ja,Valiviita:2009nu,Majerotto:2009np,Gavela:2010tm,Honorez:2010rr,Martinelli:2010rt,LopezHonorez:2010ij}. Coupled cosmologies, in order to satisfy
CMB constraints, predict values for the cosmological parameters today which may differ substantially from the parameters values within non-interacting cosmologies. Therefore interacting cosmologies can hide their effects at low redshifts and weak lensing measurements can help enormously in constraining dark sector coupled models.

We review here the future CMB constraints on interacting dark matter-dark energy models presented in Ref.~\cite{Martinelli:2010rt}, adding weak lensing data from the future Euclid~\cite{Refregier:2006vt,Refregier:2010ss} and LSST \cite{:2009pq}-like surveys.
Weak lensing probes are shown to be highly complementary to CMB measurements and extremely powerful tools to constrain interacting dark sector models.

The structure
of the paper is as follows. Section~\ref{sec:seci} presents the basics of coupled cosmologies. In Secs.~\ref{sec:ii} and \ref{sec:seciii} we describe the lensing
 extraction methods for galaxy surveys and CMB measurements respectively. Section~\ref{sec:iv} contains the description of the future data used in our analyses. The
results from our Markov Chain Monte Carlo (MCMC) analyses are presented in Sec.~\ref{sec:secv}. We draw our conclusions in Sec.~\ref{sec:secvi}.

\section{Coupled cosmologies}
\label{sec:seci}
At the level of the stress-energy tensor it is always
possible to introduce an interaction between the fluids of the dark sector in the following way~\cite{Kodama:1985bj}:
\begin{eqnarray}
\nabla_\mu T^\mu_{({\rm dm})\nu} =Q_\nu \quad\mbox{and}\quad
\nabla_\mu T^\mu_{({\rm de})\nu} =-Q_\nu.
\label{eq:conservDMDE}
\end{eqnarray}
The  4-vector $Q_\nu$ governs the energy-momentum transfer between the dark
components and  $T^\mu_{({\rm dm})\nu}$ and $T^\mu_{({\rm de})\nu}$
are the energy-momentum tensors for the dark matter and dark energy fluids, respectively. The momentum exchange  $Q_\nu$ can be parallel to the dark energy four
 velocity $u_{\nu}^{({\rm de})}$ or to the dark matter four velocity $u_{\nu}^{({\rm dm})}$. The first option include  all quintessence coupled models and are
 effectively ``modified gravity'' models, implying the presence of a ``fifth force'' effect (only for the dark matter), that is, a violation of the equivalence
principle. For both options the evolution equations for the dark matter and dark energy background energy densities are identical and reduce to:
\begin{eqnarray}
   \label{eq:EOMm}
  \dot{\bar \rho}_{\rm dm}+ 3  \mathcal{H} \bar\rho_{\rm dm} &=&a \bar Q\,,\\
\label{eq:EOMe}
 \dot{\bar \rho}_{\rm de}+ 3 \mathcal{H} \bar \rho_{\rm de}(1+ w)&=&-a \bar Q\,,
\end{eqnarray}
where the bars denote background quantities, the dot indicates derivative with respect to conformal time $d\tau = dt/a$, $\mathcal{H}= {\dot a}/a$ and  $w$ is the
dark-energy equation of state. For $Q>0$, the energy flows from the dark energy system to dark matter one. For $Q<0$, the energy flow is reversed. In coupled
cosmologies the momentum exchange can be proportional to the dark matter energy density ($Q\propto \rho_{\rm dm}$) or proportional to the dark energy energy density ($Q\propto \rho_{\rm de}$).
. However, even if models proportional to the dark matter and dark energy velocities provide the same background history, the perturbation evolution is dramatically
different. Therefore, while geometrical probes alone are unable to distinguish among the two of them, probes of the perturbation evolution via weak lensing measurements will make
these two models fundamentally different. Another aspect of coupled models is that they can show non adiabatic, early time
instabilities~\cite{Valiviita:2008iv,He:2008si,Chongchitnan:2008ry,Gavela:2009cy,Jackson:2009mz,Corasaniti:2008kx,Majerotto:2009np} due to the dark coupling term
 which appears in the dark energy pressure perturbations. In the following, we shall restrict our analyses to coupled models which satisfy the stability criterion of Ref.~\cite{Gavela:2009cy} and therefore are free of early-time, non adiabatic instabilities. We consider the dark coupled model of Ref.~\cite{Gavela:2009cy} (see also Ref.~\cite{Gavela:2010tm} for the perturbation analysis details)
\begin{equation}
  Q_\nu^{({\rm dm})}= \xi H \rho_{\rm de} u_{\nu}^{({\rm dm})}=-Q_\nu^{({\rm de})} ~,
\label{eq:coupl}
\end{equation}
where $\xi$ is a dimensionless coupling (considered constant, as well
as the dark energy equation of state $w$, in the present analysis). $H$ and $\rho_{\rm de}$ refer to the total
expansion rate and dark energy density, background plus perturbation,
{\it i.e} $H={\cal H}/a+\delta H$ and $\rho_{\rm de}=\bar
\rho_{\rm de}+\delta \rho_{\rm de}$ respectively.
Notice from Eq.~(\ref{eq:coupl}) that $Q_\nu^{({\rm a})}$ has been chosen parallel
to the dark matter four velocity $u_{\nu}^{({\rm dm})}$, in order to avoid
momentum transfer in the rest frame of the dark
matter component~\cite{Valiviita:2008iv}. For this choice of energy exchange
$Q_\nu^{({\rm a})}$, positive (negative) values of the coupling $\xi$ will lead to
lower (higher) dark matter energy densities in the past than in the
uncoupled $\xi=0$ case. We only consider here negative couplings and $w > -1$,
 avoiding the instability problems previously mentioned, see Ref.~\cite{Gavela:2009cy} for details. For the numerical analyses presented here,
 we have modified the publicly available CAMB code~\cite{camb}, taking into account the presence of the dark coupling in both the background and the linear
perturbation equations.

\section{Galaxy weak Lensing} \label{sec:ii}
Weak gravitational lensing of the images of distant galaxies offers a useful geometrical way
to map the matter distribution in the Universe.
Following Ref. \cite{Bartelmann:1999yn} one can describe the distortion of the images of distant galaxies through the tensor:\\
\begin{center}
\(\psi_{ij}=\left(
\begin{array}{cc}
-\kappa-\gamma_1 & -\gamma_2 \\
-\gamma_2 & -\kappa+\gamma_2 \\
\end{array}\right)\)
\end{center}

where $\kappa$ is the convergence term and $\gamma=\gamma_1+i\gamma_2$ is the complex shear field.
As shown in Ref. \cite{Huterer:2010hw} the shear and the convergence terms can be written as a function of the projected Newtonian potentials $\psi_{,ij}$:
$$\gamma=\frac{1}{2}(\psi_{,11}-\psi_{,22})+i\psi_{,12} \ ,$$
$$k=\frac{1}{2}(\psi_{,11}-\psi_{,22})$$
where the commas indicate the derivatives with respect to the directions transverse to the line of sight and the projected potentials are
$\psi_{,ij}=-(1/2)\int g(z)(\Psi_{,ij}+\Phi_{,ij})dz$ with the lensing kernel :
$$g(z)=\int dz^\prime\frac{n(z^\prime)D_A(z,z^\prime)}{D_A(0,z^\prime)} \ .$$
Here $n(z)$ is the galaxy redshift distribution. In our analysis we assume flatness of the Universe. However in general the angular diameter distance $D_A$ between
the lens and the source depends on the spatial curvature $K$: $$D_A=\frac{1}{\sqrt{K}}\sin(\sqrt{K}r),\phantom{aaa}K>0$$ $$D_A=r,
\phantom{aaa}K=0$$ $$D_A=\frac{1}{\sqrt{-K}}\sinh(\sqrt{-K}r),\phantom{aaa} K<0$$ and the comoving distance is:
$$r(z,z^\prime)=\int_z^{z^\prime}\frac{dz^\prime}{E(z^\prime)}$$ with $E(z)=H(z)/H_0$.

Images distortions induced by the matter distribution are generally
small. To extract cosmological information it is hence necessary to statistically analyze a large number of images. The two point correlation function of the
convergence is at present the best measured statistic of the weak lensing but, of course, also higher order statistics contains cosmological information. It is
convenient to work in the multipole space and define the convergence power spectrum as the harmonic transform of the two-point correlation function. This is usually
the most analyzed and studied statistical quantity related to the weak lensing and we will focus on the convergence power spectra in order to properly compare our
results to similar analysis in literature. However it should be stressed that, as shown in \cite{Schneider:2002jd}, the convergence power spectrum is only indirectly
and partially obtainable from the two point correlation function. 

Future surveys will measure redshifts of
billions of galaxies allowing the possibility of a tomographic reconstruction of the matter distribution. We can define hence the convergence power spectra in each
redshift bin and the cross-power spectra:
\begin{equation}\label{pidielle}
 P_{jk}(\ell)=H_0^3\int_0^\infty\frac{dz}{E(z)}W_i(z)W_j(z)P_{NL}[P_L\left(\frac{H_0\ell}{r(z)},z\right)]
\end{equation}
where $P_{NL}$ is the non-linear matter power spectrum at redshift
$z$, obtained correcting the linear one $P_{L}$. $W(z)$ is a weighting function:
\begin{equation}\label{weight}
    W_i(z)=\frac{3}{2}\Omega_m(1+z)\int_{z_i}^{z_{i+1}}dz^\prime\frac{n_i(z^\prime)r(z,z^\prime)}{r(0,z^\prime)}
\end{equation}
with subscripts $i$ and $j$ indicating the redshift bin.
Equation (\ref{pidielle}) shows the cosmological information contained in weak lensing measurements: the function $W(z)$ encodes the
information on how the three-dimensional matter distribution is projected on the sky, while the matter power spectrum quantifies the overall matter distribution.\\
The observed convergence power
spectra is affected mainly by systematic uncertainties arising from the
intrinsic ellipticity of galaxies $\gamma^2_{rms}$. These uncertainties can
be reduced averaging over a large number of sources. The observed
convergence power spectra will be hence:
\begin{equation}\label{obsconv}
    C_{jk}=P_{jk}+\delta_{jk}\gamma^2_{rms}\tilde{n}_j^{-1}
\end{equation}
where $\tilde{n}_j$ is the number of sources per steradian in the
$j-th$ bin.

\section{CMB Lensing extraction}
\label{sec:seciii}
The analysis presented here includes, in addition to the primary CMB anisotropy angular power spectrum, the information from CMB lensing. Gravitational CMB lensing,
as already shown (see e.g. \cite{Perotto:2006rj,Calabrese:2009tt}) can improve significantly the CMB constraints on several cosmological parameters, since it is
strongly connected with the growth of perturbations and gravitational
potentials at redshifts $z < 1$ and therefore, it can break important
degeneracies.
The lensing deflection field $d$ can be related to the lensing potential $\phi$ as $d=\nabla\phi$~\cite{hirata:2003}. In harmonic space, the deflection and lensing potential multipoles follow
\begin{equation}
d_l^m=-i\sqrt{l(l+1)}\phi_l^m,
\label{eq:defd}
\end{equation}
and therefore, the power spectra $C^{dd}_l\equiv\left\langle d_l^m d_l^{m*}\right\rangle$ and
$C_l^{\phi\phi}\equiv\left\langle\phi_l^m\phi_l^{m*}\right\rangle$ are related through
\begin{equation}
C_l^{dd}=l(l+1)C_l^{\phi\phi}.
\end{equation}
Lensing introduces a correlation between different CMB multipoles (that otherwise would be fully uncorrelated) through the relation
\begin{equation}
\left\langle a_l^m b_{l'}^{m'}\right\rangle=(-1)^m\delta_m^{m'}\delta_l^{l'}C_l^{ab}+
\sum_{LM}{\Xi^{mm'M}_{l\ l'\ L}\phi^M_L}~,
\label{eq:lens_corr}
\end{equation}
where $a$ and $b$ are the ${T,E,B}$ modes and $\Xi$ is a linear combination of the unlensed
power spectra $\tilde{C}_l^{ab}$ (see \cite{lensextr} for details).\\
In order to obtain the deflection power spectrum from the observed $C_l^{ab}$, we have to invert Eq.~(\ref{eq:lens_corr}), defining a quadratic estimator for the
deflection field given by
\begin{equation}
d(a,b)_L^M=n_L^{ab}\sum_{ll'mm'}W(a,b)_{l\ l'\ L}^{mm'M}a^m_lb^{m'}_{l'}~,
\label{eq:estimator}
\end{equation}
where $n_L^{ab}$ is a normalization factor needed to construct an unbiased estimator ($d(a,b)$ must satisfy Eq.~(\ref{eq:defd})).
The variance of this estimator reads
\begin{equation}
 \langle d(a,b)_L^{M*} d(a',b')_{L'}^{M'}\rangle\equiv \delta_{L}^{L'}\delta^{M'}_{M}(C_L^{dd}+N_L^{aa'bb'})~,
\end{equation}
and depends on the choice of the weighting factor $W$
and leads to a noise $N_L^{aa'bb'}$ on the deflection power spectrum $C_L^{dd}$ obtained through this method. In the next section we  describe the method followed here to extract the lensing noise. \\

\section{Future data analysis} \label{sec:iv}

\subsection{Galaxy weak lensing data}
Future weak lensing surveys will measure photometric redshifts of
billions of galaxies allowing the possibility of 3D weak lensing
analysis
(e.g.\cite{Heavens:2003,Castro:2005,Heavens:2006,Kitching:2007}) or
a tomographic reconstruction of growth of structure as a function
of time through a binning of the redshift distribution of galaxies,
with a considerable gain of cosmological information (e.g. on
neutrinos \cite{Hannestad:2006as}; dark energy \cite{Kitching:2007};
the growth of structure \cite{art:Bacon,art:Massey}  and the
dark matter distribution as a function of redshift
\cite{art:Taylor}).

Here we use the typical specifications for future weak lensing surveys like those of the Euclid and LSST experiments. Euclid will observe about $35$ galaxies per square
arcminute in the redshift range $0.5<z<2$ with an uncertainty of about $\sigma_z=0.05(1+z)$ (see \cite{Refregier:2006vt}).
LSST is expected to have similar characteristics, with slightly higher number density of sources and larger redshift range, but also with a higher intrinsic shear.  We build mock datasets of convergence power spectra for these two surveys. Tables \ref{tabeuclid}  and \ref{tablsst} show the number of galaxies per
arcminute$ ^{-2}$ ($n_{gal}$), redshift range, $f_{sky}$ and
intrinsic ellipticity for these surveys.
\begin{table}[h]
\begin{center}
\begin{tabular}{cccccccc}
$n_{gal} (arcmin^{-2})$ \hspace{5pt} & redshift\hspace{5pt}  &$f_{sky}$\hspace{5pt}  & $\gamma_{rms}$\\
\hline &&&&&&\\
$35$\hspace{5pt}&$0.5<z<2$\hspace{5pt} &$0.5$\hspace{5pt} &$0.22$\hspace{5pt}\\
\hline \hline
\end{tabular}
\caption{Specifications for the Euclid like survey considered in
this paper. The table shows the number of galaxies per square
arcminute ($n_{gal}$), redshift range, $f_{sky}$ and intrinsic ellipticity
($\gamma^2_{rms}$) per component.} \label{tabeuclid}
\end{center}
\end{table}

\begin{table}[h]
\begin{center}
\begin{tabular}{cccccccc}
$n_{gal} (arcmin^{-2})$ \hspace{5pt} & redshift\hspace{5pt}  &$f_{sky}$\hspace{5pt}  & $\gamma_{rms}$\\
\hline &&&&&&\\
$40$\hspace{5pt}&$0<z<3$\hspace{5pt} &$0.5$\hspace{5pt} &$0.28$\hspace{5pt}\\
\hline \hline
\end{tabular}
\caption{Specifications for the LSST like survey considered in
this paper. The table shows the number of galaxies per square
arcminute ($n_{gal}$), redshift range, $f_{sky}$ and intrinsic ellipticity dispersion
($\gamma^2_{rms}$) per component.} \label{tablsst}
\end{center}
\end{table}

 The expected $1\sigma$ uncertainty
on the convergence power spectra $P(\ell)$ is given by
\cite{Cooray:1999rv}:
\begin{equation}\label{sigmaconv}
    \sigma_{\ell}=\sqrt{\frac{2}{(2\ell+1)f_{sky}\Delta_{\ell}}}\left(P(\ell)+\frac{\gamma_{rms}^2}{n_{gal}}\right)~,
\end{equation}
where $\Delta_{\ell}$ is the $\ell$-bin width used to generate data. Here we choose
$\Delta_\ell=1$ for the range $2<\ell<100$ and $\Delta_\ell=40$ for
$100<\ell<1500$. For the convergence power spectra we use $\ell_{max}=1500$ in order
to exclude the scales where the non-linear growth of structure is more relevant
and the shape of the non-linear matter power spectra is, as a
consequence, more uncertain (see \cite{Smith:2002dz}). We describe the galaxy distribution of Euclid survey as
in \cite{Abdalla:2007uc}, $n(z)\propto z^2\exp(-(z/z_0)^{1.5})$ where $z_0$ is set by the median redshift of the sources, $z_0=z_m/1.41$. Here we calculate
the power spectra assuming a median redshift $z_m=1$.  Although this assumption is reasonable for the Euclid survey, it is known that the parameters that control the shape of the distribution function may have strong degeneracies with some cosmological parameters as the matter density, $\sigma_8$ and the spectral index~\cite{Fu:2007qq}. \\

\subsection{CMB data}

We create a full mock CMB dataset (temperature, E--polarization
mode and lensing deflection field) with noise properties consistent
with the Planck~\cite{:2006uk} experiment (see Tab.~\ref{tab:exp}
for specifications).

\begin{table}[!htb]
\begin{center}
\begin{tabular}{rccc}
Experiment & Channel & FWHM & $\Delta T/T$ \\
\hline
Planck & 70 & 14' & 4.7\\
\phantom{Planck} & 100 & 10' & 2.5\\
\phantom{Planck} & 143 & 7.1'& 2.2\\

$f_{sky}=0.85$ & & & \\
\hline
\hline
\end{tabular}
\caption{Planck experimental specifications. Channel frequency is given in GHz, FWHM (Full-Width at Half-Maximum) in arc-minutes, and the temperature
sensitivity per pixel in $\mu K/K$. The polarization sensitivity is $\Delta E/E=\Delta B/B= \sqrt{2}\Delta T/T$.}
\label{tab:exp}
\end{center}
\end{table}

We consider for each channel a detector noise of $w^{-1} =
(\theta\sigma)^2$, where $\theta$ is the FWHM (Full-Width at
Half-Maximum) of the beam assuming a Gaussian profile and $\sigma$ is
the temperature sensitivity $\Delta T$ (see Tab.~\ref{tab:exp} for the polarization sensitivity). We therefore add to each $C_\ell$ fiducial spectra a
noise spectrum given by:
\begin{equation}
N_\ell = w^{-1}\exp(\ell(\ell+1)/\ell_b^2) \, ,
\end{equation}
where $\ell_b$ is given by $\ell_b \equiv \sqrt{8\ln2}/\theta$.

We make use of the method presented in \cite{lensextr} to
construct the weighting factor $W$ of Eq.~(\ref{eq:estimator}). In
that paper, the authors choose $W$ to be a function of the power
spectra $C_\ell^{ab}$, which include both CMB lensing and primary
anisotropy contributions. This choice leads to five quadratic
estimators, with $ab={TT,TE,EE,EB,TB}$; the $BB$ case is excluded
because the method of Ref.~\cite{lensextr} is only valid when the
lensing contribution is negligible compared to the primary
anisotropy, assumption that fails for the B modes in the case of Planck. \\
The five quadratic estimators can be combined into a minimum
variance estimator which provides the noise on the deflection
field power spectrum $C_\ell^{dd}$:
\begin{equation}
N_\ell^{dd}=\frac{1}{\sum_{aa'bb'}{(N_\ell^{aba'b'})^{-1}}}~.
\end{equation}
We compute the minimum variance lensing noise for the Planck experiment by means of a routine publicly available at \lensnoise.
The datasets (which include the lensing deflection power spectrum) are analyzed with a full-sky exact likelihood routine available at the same URL.

\subsection{Analysis method}

We perform two different analyses. First, we compute the expected constraints on the coupling parameter $\xi$ from Planck and Euclid data, comparing the results with the limits arising from Planck and LSST data. Secondly, we investigate the effects of a wrong assumption about the interaction between dark matter and dark energy on the values of the cosmological parameters: we generate a dataset with a non-zero $\xi$ fiducial value but analyze the data assuming that there is no coupling between the dark components ($\xi=0$). We perform a MCMC analysis  based on the publicly available
package \texttt{cosmomc} \cite{Lewis:2002ah} with a convergence
diagnostic using the Gelman and Rubin statistics. 

We sample the following  set of cosmological parameters, adopting flat priors on
them: the baryon and cold dark matter densities $\Omega_{b}h^2$ and
$\Omega_{c}h^2$, the ratio of the sound horizon to the angular
diameter distance at decoupling $\theta_s$, the scalar spectral
index $n_s$, the overall normalization of the spectrum $A_s$ at
$k=0.002$ {\rm Mpc}$^{-1}$, the optical depth to reionization
$\tau$, and, finally, the coupling parameter $\xi$.

The fiducial model for the standard cosmological parameters
is the best-fit from the WMAP seven year data analysis of
Ref.~\cite{wmap7} with $\Omega_{b}h^2=0.02258$, $\Omega_{c}h^2=
0.1109$, $n_s=0.963$, $\tau=0.088$, $A_s=2.43\times10^{-9}$ and
$\Theta=1.0388$. For the coupling parameter we first assume a fiducial value $\xi=0$
to test the constraints achievable on the coupling model.
Finally, we analyse a dataset with a fiducial value
$\xi=-0.1$ assuming (wrongly) a $\Lambda$CDM scenario with
 $\xi=0$, with Planck and Euclid forecasted data.
This exercise will allow us to investigate the bias introduced on
 cosmological parameter inference from a wrong assumption about the
coupling model.

\section{Results}
\label{sec:secv}

\begin{figure}[h!]
\begin{center}
\hspace*{-1cm}
\begin{tabular}{cc}
\includegraphics[width=8cm]{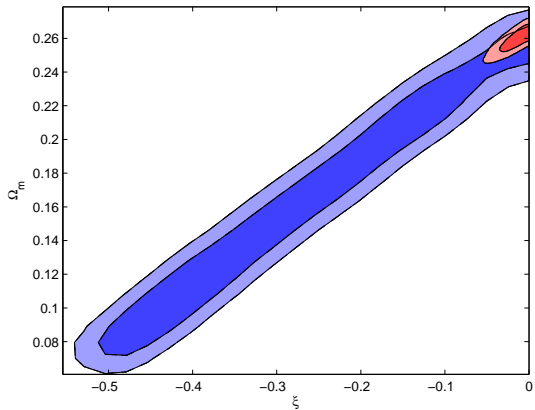}&
 \end{tabular}
\caption{2-D constraints on $\xi$ and $\Omega_m$ using Planck (blue contours) and Planck plus Euclid (red contours).}
\label{fig:compare}
\end{center}
\end{figure}

\begin{figure}
\begin{center}
\hspace*{-1cm}
\begin{tabular}{cc}
\includegraphics[width=8cm]{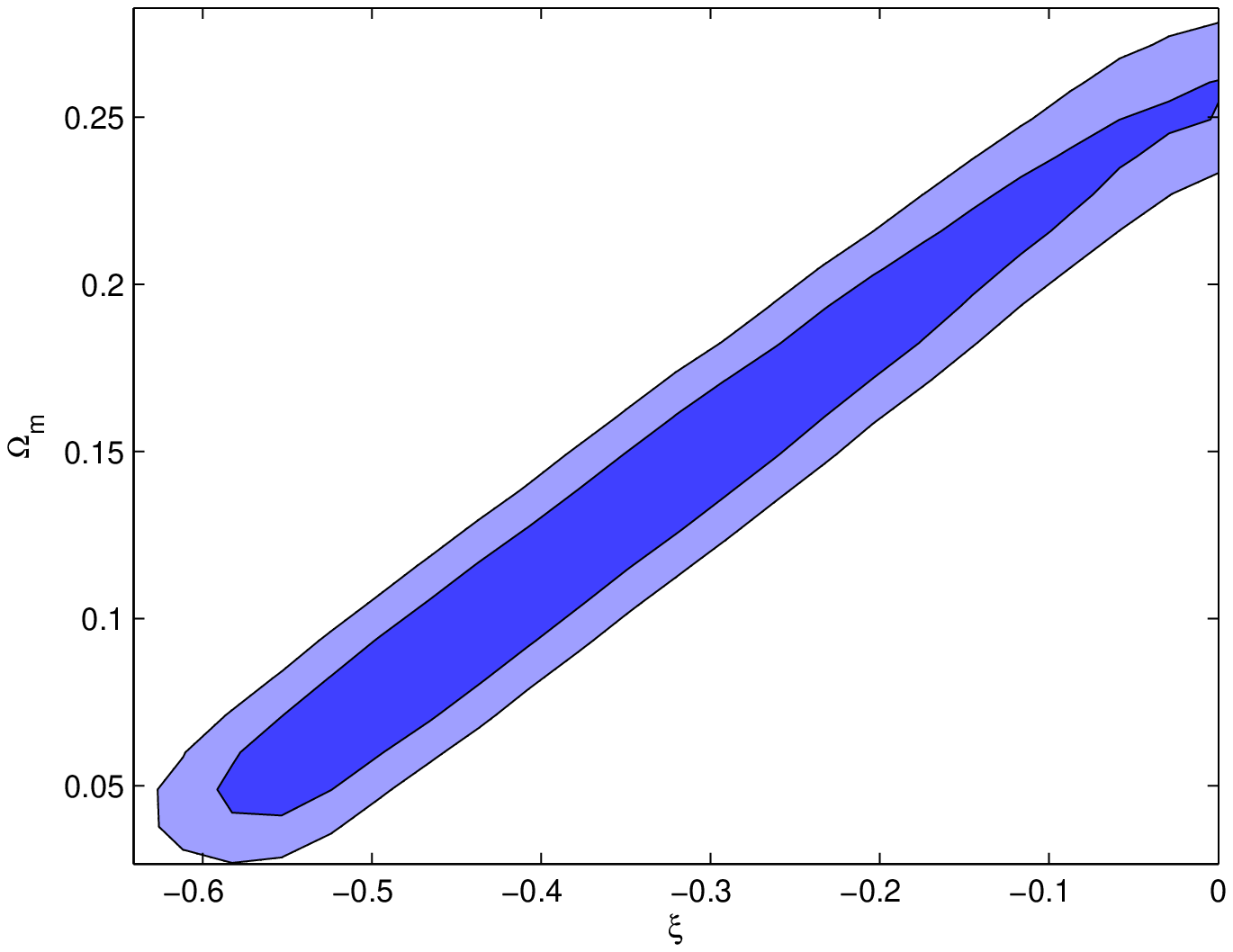}\\
\includegraphics[width=8cm]{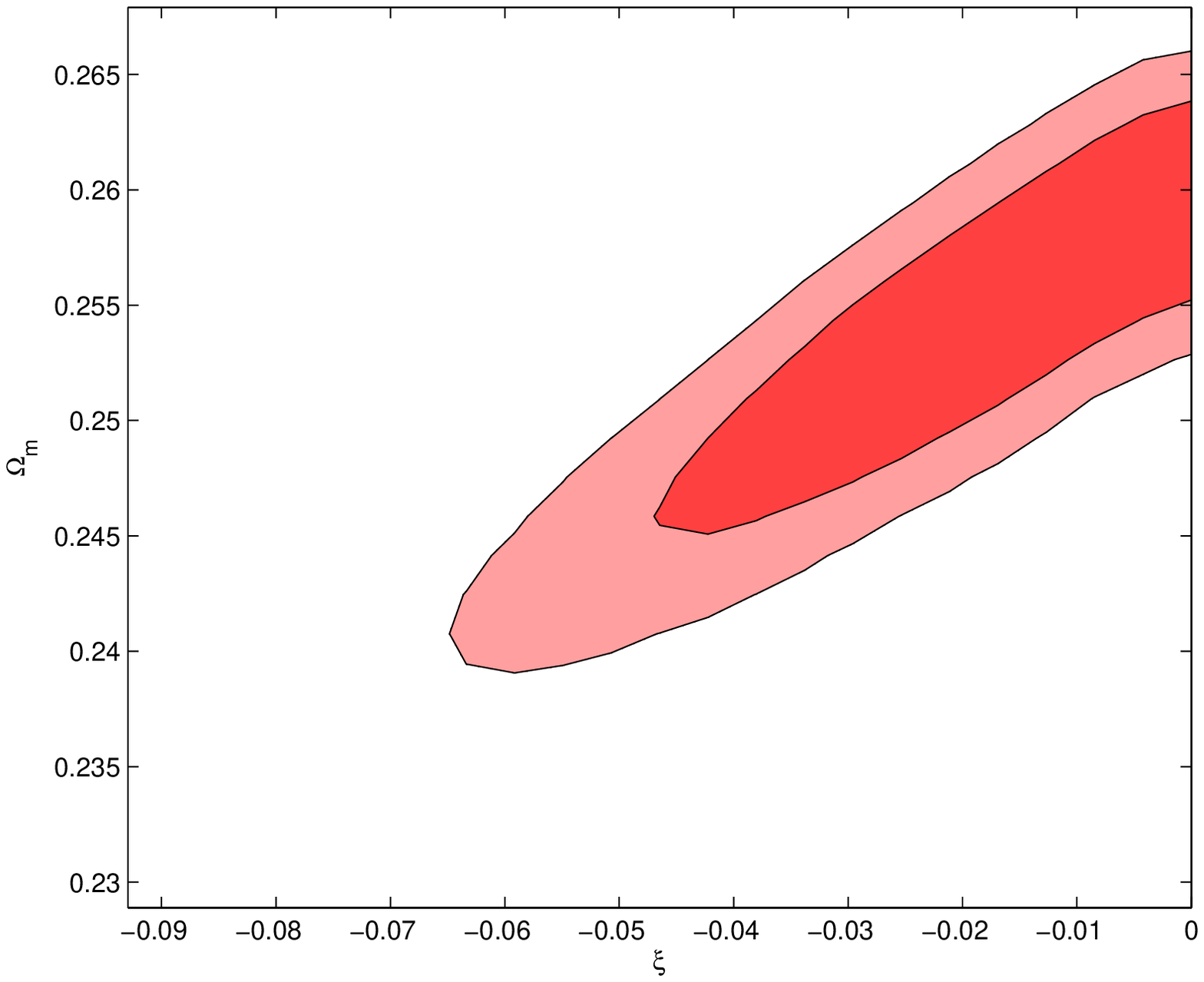}\\
 \end{tabular}
\caption{Top panel: $68\%$ and $95 \%$ c.l. contours in the ($\Omega_m$, $\xi$) plane from Planck data only. Bottom panel: same as in the top panel, but for the
combination of Planck plus Euclid data (note the different scale for the x-axis).}\label{fig:countours1}
\end{center}
\end{figure}

\begin{figure}
\begin{center}
\hspace*{-1cm}
\begin{tabular}{cc}
\includegraphics[width=8cm]{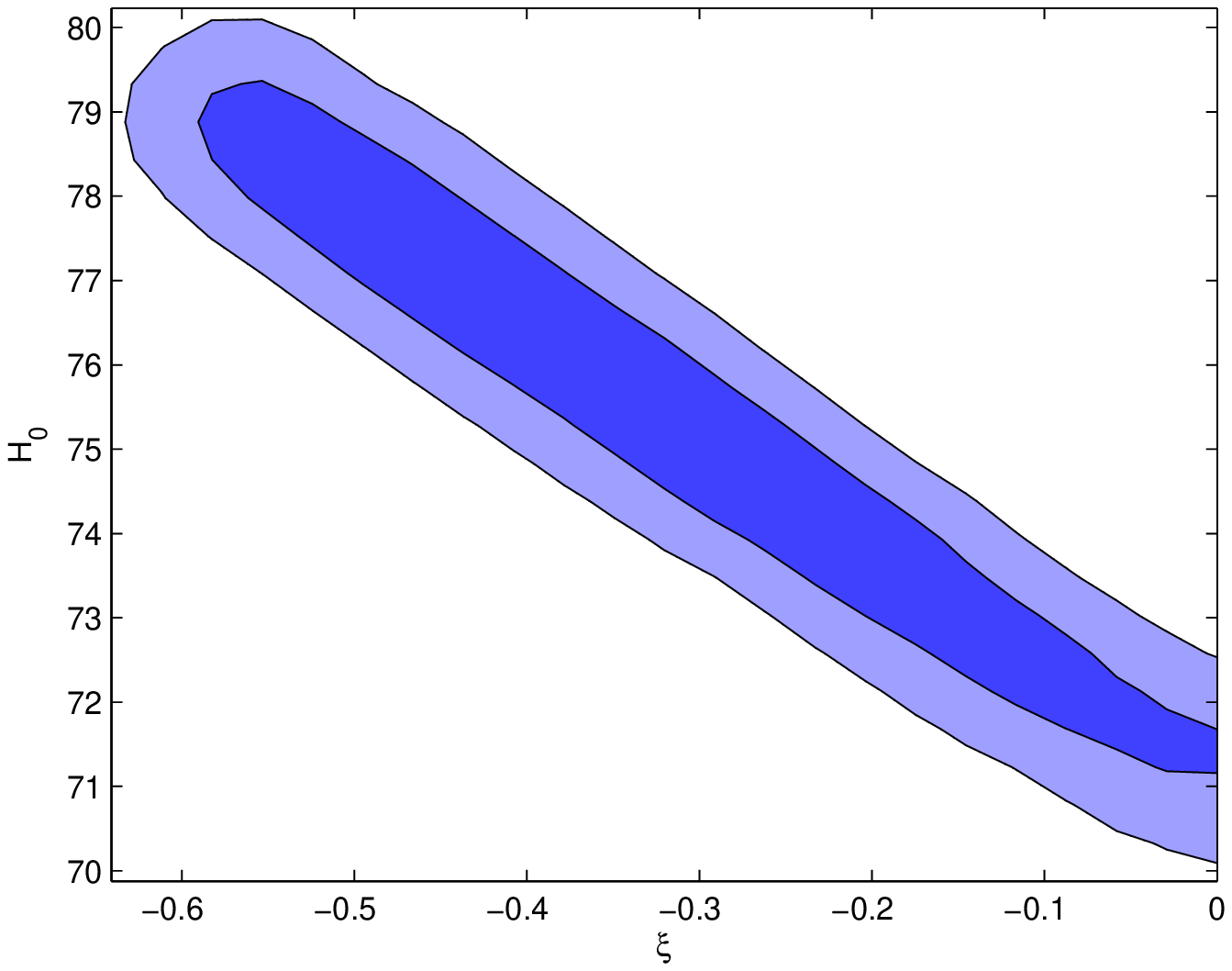}\\
\includegraphics[width=8cm]{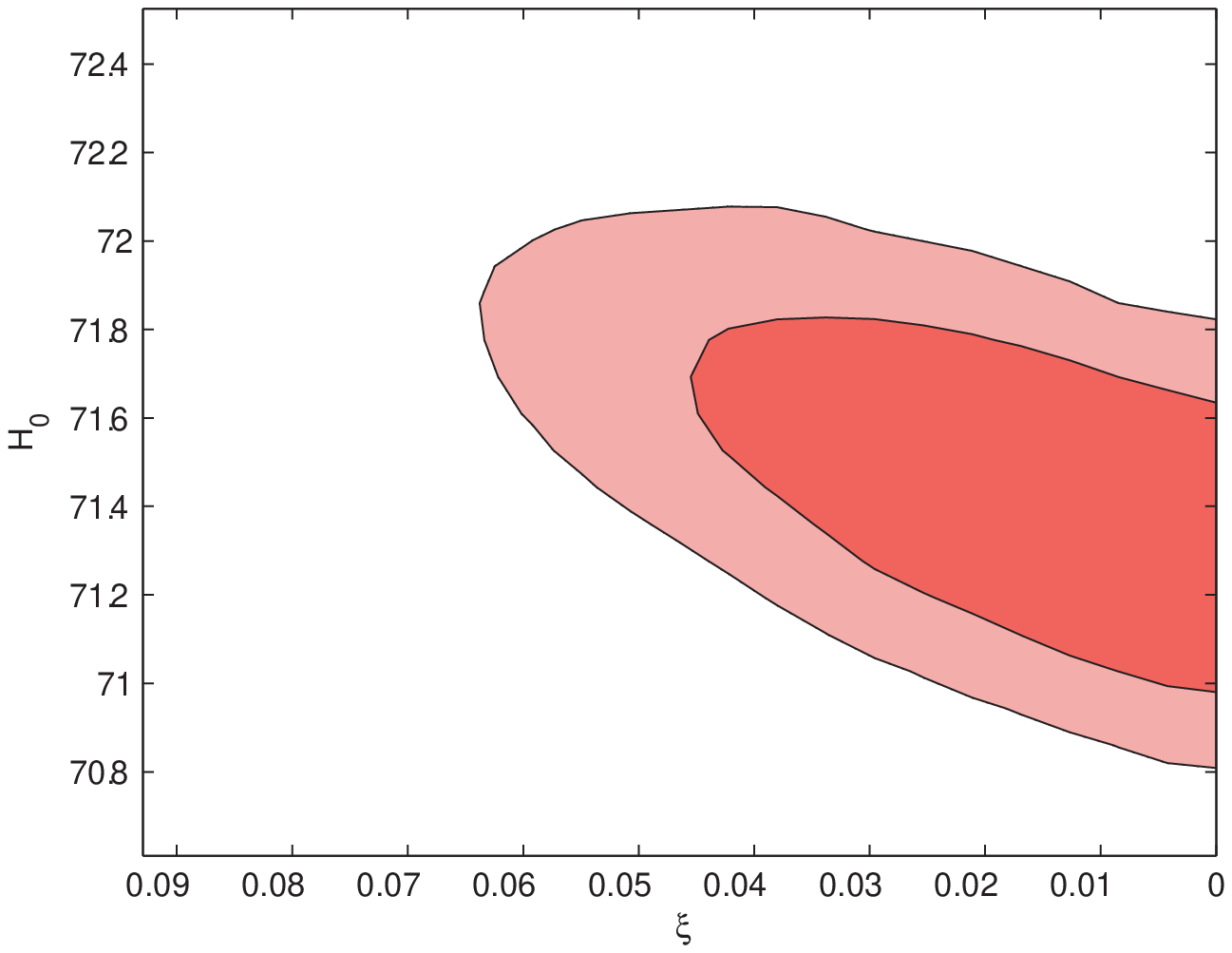}\\
 \end{tabular}
\caption{Top panel: $68\%$ and $95 \%$ c.l. contours in the ($H_0$, $\xi$) plane from Planck data only. Bottom panel: same as in the top panel, but for the
combination of Planck plus Euclid data (note the different scale for the x-axis).}\label{fig:countours2}
\end{center}
\end{figure}

In Table \ref{tab:results} we show the MCMC constraints at $68 \%$ c.l. for the coupled universe
from Planck data alone and from Planck data combined with Euclid data.
For this last case we also fit the data fixing $\xi$ to $0$,
thus performing a standard analysis in a universe where dark matter and dark energy are not interacting,
in order to show the importance of the degeneracies introduced by
the presence of a coupling $\xi$ on the other cosmological parameters errors.
There is a very high level of correlation among the dimensionless coupling $\xi$ and the parameters $H_0$ and $\Omega_ch^2$ in the Planck analysis
(see also Figs.~\ref{fig:countours1} and \ref{fig:countours2}). When Planck and Euclid data are combined, the degeneracy between $\xi$ and $H_0$ is broken,
leading to a much better constrain on the coupling parameter $\xi$ than when using CMB data alone\cite{Martinelli:2010rt}, as one can notice from Table
\ref{tab:results} and Fig.~\ref{fig:compare}. However, the degeneracy between $\xi$ and $\Omega_ch^2$ is not broken by the combination of Planck and Euclid, thus it will be possible to
further improve the constraints on $\xi$ with independent measurements of $\Omega_c$. We also note that the constraints on the standard cosmological parameters
are in good agreement with those reported in \cite{Refregier:2010ss}.

\begin{table}[!htb]
\begin{center}
\begin{tabular}{|l|c|c|c|c|}
\hline

               & \multicolumn{2}{c|}{Planck}& \multicolumn{2}{c|}{Planck+Euclid} \\
\hline
Model & Varying $\xi$ & $\xi=0$ & Varying $\xi$ & $\xi=0$\\
Parameter &&&&\\
\hline
$\Delta{(\Omega_bh^2)}$       & $0.00013$  & $0.00013$    & $0.00010$    &  $0.00010$ \\
$\Delta{(\Omega_ch^2)}$       & $0.0299$   & $0.0010$     & $0.0024$     &  $0.00055$ \\
$\Delta{(\theta_s)}$          & $0.0023$   & $0.00026$    & $0.00027$    &  $0.00023$ \\
$\Delta{(\tau)}$              & $0.0042$   & $0.0042$     & $0.0026$     &  $0.0026$  \\
$\Delta{(n_s)}$               & $0.0030$   & $0.0031$     & $0.0029$     &  $0.0027$  \\
$\Delta{(\log[10^{10} A_s])}$ & $0.013$    & $0.013$      & $0.0097$     &  $0.0093$  \\
$\Delta{(H_0)}$               & $2.28$     & $0.43$       & $0.29$       &  $0.27$    \\
$\Delta{(\Omega_\Lambda)}$    & $0.0614$   & $0.0050$     & $0.0062$     &  $0.0026$  \\
$\xi$                         & $>-0.56$   & $-$          & $>-0.05$     &  $-$       \\
\hline
\end{tabular}
\caption{$68 \%$ c.l. errors on cosmological parameters. Upper
limits on $\xi$ are $95\%$ c.l. constraints.}
\label{tab:results}
\end{center}
\end{table}

As stated above, future surveys like Euclid will be able to tomographically reconstruct the matter distribution. Exploiting this possibility would
improve the constraints, but, as already pointed out in \cite{Martinelli:2010wn}, the non tomographic analysis can be thought as a conservative estimation
of the constraints as we are not including systematic effects.\\

\begin{figure}[h!]
\begin{center}
\hspace*{-1cm}
\begin{tabular}{cc}
\includegraphics[width=8cm]{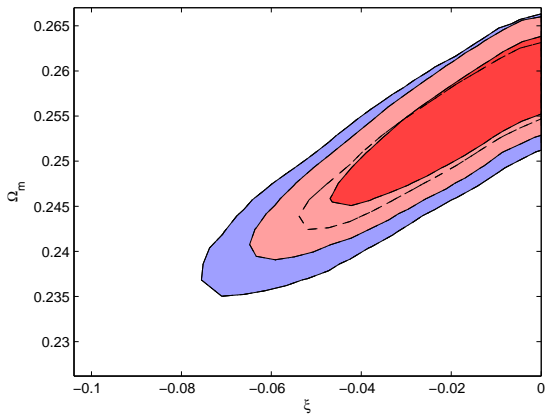}&
 \end{tabular}
\caption{2-D constraints on $\xi$ and $\Omega_m$ using Planck and LSST data (blue contours) and Planck and Euclid data (red contours).}
\label{fig:lsst}
\end{center}
\end{figure}

Table \ref{tab:lsst} contains both the results from the combination of Planck and Euclid data and those from the combination of Planck and LSST. Notice that the
results are quite similar. However, the slightly better constraints on $\xi$ from the Planck plus Euclid combination leads to a better measurement of the cold dark
 matter content of the universe than the one performed by Planck plus LSST data (see also Fig.~\ref{fig:lsst}).\\
Moreover, the difference between Planck+Euclid and Planck+LSST results would be bigger if systematic effects are included, as LSST, being a ground based survey,
will be more affected by these.\\

\begin{table}[!htb]
\begin{center}
\begin{tabular}{|l|c|c|}
\hline

               & Planck+Euclid& Planck+LSST \\
\hline
Parameter &&\\
\hline
$\Delta{(\Omega_bh^2)}$       & $0.00010$    &  $0.00010$ \\
$\Delta{(\Omega_ch^2)}$       & $0.0024$     &  $0.0026$ \\
$\Delta{(\theta_s)}$          & $0.00027$    &  $0.00028$ \\
$\Delta{(\tau)}$              & $0.0026$     &  $0.0027$  \\
$\Delta{(n_s)}$               & $0.0029$     &  $0.0029$  \\
$\Delta{(\log[10^{10} A_s])}$ & $0.0097$      &  $0.010$  \\
$\Delta{(H_0)}$               & $0.29$       &  $0.29$    \\
$\Delta{(\Omega_\Lambda)}$    & $0.0062$     &  $0.0065$  \\
$\xi$                         & $>-0.04$     &  $>-0.06$       \\
\hline
\end{tabular}
\caption{$68 \%$ c.l. errors on cosmological parameters. Upper
limits on $\xi$ are $95\%$ c.l. constraints.}
\label{tab:lsst}
\end{center}
\end{table}

\begin{figure}[h!]
\begin{center}
\hspace*{-1cm}
\includegraphics[width=8cm]{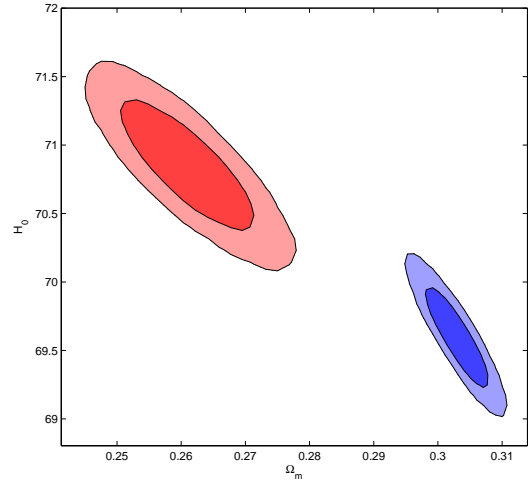}
\caption{$68\%$ and $95 \%$ confidence levels from Planck plus Euclid data when a fiducial cosmology with $\xi=-0.1$ is fitted to a non interacting cosmology
with $\xi$ fixed to $0$ (blue countours) or to an interacting cosmology in which $\xi$ is allowed to vary.}\label{fig:shift}
\end{center}
\end{figure}

In addition, we have also (wrongly) fitted a mock dataset with
$\xi=-0.1$ to a non interacting cosmology in which the dimensionless coupling vanishes ($\xi=0$). From this exercise we
find a consistent bias in the recovered best fit value of the
cosmological parameters due to the strong degeneracies among
$\xi$ and both the Hubble constant $H_0$ and the matter energy density $\Omega_m$ parameters, see Tab.\ref{tab:shift}.
Note, from the results depicted in Figs. \ref{fig:countours1}, \ref{fig:countours2} and \ref{fig:shift} and also from the results in Tab.~\ref{tab:shift} that the
shift in the best fit values is, as expected, along the direction of the degeneracy of $\xi$ with these parameters. These results show that even
for a small value of $\xi$, the best fit values recovered by wrongly assuming that t        here is no dark coupling are more than $68 \%$ c.l. (for
some parameters at more than $95 \%$ c.l.) away from the correct fiducial
values, and may induce an underestimation of both $H_0$ and $\sigma_8$
and an overestimation of $\Omega_ch^2$. In the last column in Tab.~\ref{tab:shift} we show the difference between the \textit{wrong} value estimated fixing $\xi=0$ and the fiducial value, relative to the 1$\sigma$ error: as expected the largest shifts are in the parameters that are directly involved in determining the energy momentum transfer between components, namely matter and dark energy densities and Hubble parameter (see section \ref{sec:seci}). We note that also other parameters, as $\sigma_8$ and $n_s$, have significant shifts.\\We conclude, hence, that future analyses of high precision data
from Euclid and Planck need to consider possible deviations from the minimal $\Lambda$CDM scenario in order to avoid biases in the measurements of the cosmological
parameters.\\

\begin{table}[!htb]\footnotesize
\begin{center}
\begin{tabular}{|l|c|c|c|}
\hline
& \multicolumn{2}{c|}{Planck+Euclid} & Fiducial values\\
\hline
Model: &$\xi=0$& varying $\xi$&\\
Parameter & & &\\
\hline
$\Omega_bh^2$       &  $0.02259\pm0.00010$    & $0.02257\pm0.00010$  & $0.02258$\\
$\Omega_ch^2$       &  $0.1245\pm0.00061$      & $0.1083\pm0.0024$    & $0.1109$\\
$\tau$              &  $0.086\pm0.0029$       & $0.090\pm0.0033$     & $0.088$\\
$n_s$               &  $0.955\pm0.0014$       & $0.961\pm0.0028$     & $0.963$\\
$H_0$               &  $69.6\pm0.23$          & $70.9\pm0.30$        & $71.0$\\
$\Omega_\Lambda$    &  $0.697\pm0.0031$       & $0.739\pm0.0066$     & $0.735$\\
$\sigma_8$          &  $0.752\pm0.00044$     & $0.841\pm0.019$     & $0.82$\\
\hline
\end{tabular}
\caption{Best fit values and $68 \%$ c.l. errors on cosmological
parameters for the case in which a fiducial model with $\xi=-0.1$ is
fitted to a $\Lambda$CDM model where $\xi=0$ is assumed. 
}
\label{tab:shift}
\end{center}
\end{table}

\section{Conclusions}
\label{sec:secvi}
The current accelerated expansion of the universe is driven by the so-called dark energy. This negative pressure component could be interpreted as the vacuum energy
density, or as a cosmic, dynamical scalar field.  If a cosmic quintessence field is present in nature, it may couple to the other fields in nature. While the couplings
 of the quintessence field to ordinary matter are severely constrained, an energy exchange among the dark matter and dark energy sectors is allowed by current
observations. Interacting cosmologies, in order to fit Cosmic Microwave Background observations, predict non standard values for the low redshift universe observables.
 Therefore, measurements of the growth of structure and of the dark matter distribution via galaxy and weak lensing surveys offer a unique window to study the nature
of the dark sectors.

The major goals of the on-going Planck experiment and the next generation of galaxy surveys are to determine the nature of the dark energy component and to measure
the remaining cosmological parameters with unprecedented precision. In this study we have exploited the capabilities of the former experiments to improve current
constraints on the dimensionless coupling $\xi$, which drives the energy flow between the dark energy and dark matter sectors~\cite{Gavela:2009cy}.

We have generated mock data for both the Planck experiment and the Euclid and LSST-like weak lensing surveys. CMB gravitational lensing extraction has also been
included in the analysis. For the weak lensing surveys, the observable we have exploited here is the convergence power spectrum. The mock data have then been analyzed
using MCMC techniques to compute the errors on the several cosmological parameters considered here. The constraints on the dimensionless coupling parameter of the
interacting model fully explored here, $\xi$, are greatly improved with respect to previous analyses in which only CMB probes were considered. We find $\xi >-0.04$ at
 the $95\%$ c.l. for the combination of Planck survey and Euclid weak lensing data. Data from a LSST-like survey combined with data from the Planck experiment could
 also provide a lower bound on $\xi$ of $-0.06$ at the $95\%$ c.l.. The coupling parameter $\xi$ is strongly degenerate with the cold dark matter density
$\Omega_{c}h^2$ and the Hubble constant $H_0$. These degeneracies may cause important biases in the cosmological parameter values if in the universe there exists an
interaction among the dark sectors. Future experiments may therefore include also coupled cosmologies as possible scenarios when fitting their data.
 Finally we conclude noting that, as recently outlined by \cite{Tarrant:2011qe}, coupled quintessence is expected to have a distinctive impact on the formation of
structures at galaxy clusters scales. In our work we limited our analysis to linear scales, but for the analysis of real future data, it will be necessary a precise
modeling of the effects of coupled quintessence on non-linear scales. At the same time, this small scale effect will offer a way to constrain coupled cosmologies
through measurements of the galaxy clusters mass distribution.

\section{Acknowledgments}
We thank Alan Heavens for very useful comments.
This work is supported by PRIN-INAF, "Astronomy probes fundamental physics".
Support was given by the Italian Space Agency through the ASI contracts "Euclid- IC" (I/031/10/0). The work of O.M. is supported by a MICINN Ram\'on y Cajal contract, by AYA2008-03531 and the Consolider Ingenio-2010 project CSD2007-00060.


\begin{thebibliography}{srt}

\bibitem{wmap7}
 E.~Komatsu {\it et al.}  [WMAP Collaboration],
  Astrophys.\ J.\ Suppl.\  {\bf 192}, 18 (2011)
  [arXiv:1001.4538 [astro-ph.CO]]~;
D.~Larson {\it et al.},
  Astrophys.\ J.\ Suppl.\  {\bf 192}, 16 (2011)
  [arXiv:1001.4635 [astro-ph.CO]].
\bibitem{sdss2}
B.~A.~Reid {\it et al.}  [SDSS Collaboration],
  Mon.\ Not.\ Roy.\ Astron.\ Soc.\  {\bf 401}, 2148 (2010)
  [arXiv:0907.1660 [astro-ph.CO]]~;
B.~A.~Reid {\it et al.},
  Mon.\ Not.\ Roy.\ Astron.\ Soc.\  {\bf 404}, 60 (2010)
  [arXiv:0907.1659 [astro-ph.CO]].
\bibitem{snalate}
R.~Amanullah {\it et al.},
  Astrophys.\ J.\  {\bf 716}, 712 (2010)
  [arXiv:1004.1711 [astro-ph.CO]].
\bibitem{Carroll:1998zi}
  S.~M.~Carroll,
  Phys.\ Rev.\ Lett.\  {\bf 81}, 3067 (1998)
  [arXiv:astro-ph/9806099].
\bibitem{Amendola:1999er}
  L.~Amendola,
  Phys.\ Rev.\  D {\bf 62}, 043511 (2000)
  [arXiv:astro-ph/9908023].
\bibitem{Amendola:1999dr}
  L.~Amendola,
  Mon.\ Not.\ Roy.\ Astron.\ Soc.\  {\bf 312}, 521 (2000)
  [arXiv:astro-ph/9906073].
\bibitem{Amendola:1999qq}
  L.~Amendola,
  Phys.\ Rev.\  D {\bf 60}, 043501 (1999)
  [arXiv:astro-ph/9904120].
\bibitem{Amendola:2000uh}
 L.~Amendola and D.~Tocchini-Valentini,
  Phys.\ Rev.\  D {\bf 64}, 043509 (2001)
  [arXiv:astro-ph/0011243].
\bibitem{Amendola:2003wa}
  L.~Amendola,
  Phys.\ Rev.\  D {\bf 69}, 103524 (2004)
  [arXiv:astro-ph/0311175].

\bibitem{Amendola:2006dg}
  L.~Amendola, G.~Camargo Campos and R.~Rosenfeld,
  Phys.\ Rev.\  D {\bf 75} (2007) 083506
  [arXiv:astro-ph/0610806].
\bibitem{Valiviita:2008iv}
  J.~Valiviita, E.~Majerotto and R.~Maartens,
  JCAP {\bf 0807}, 020 (2008)
  [arXiv:0804.0232 [astro-ph]].
\bibitem{He:2008si}
  J.~H.~He, B.~Wang and E.~Abdalla,
  Phys.\ Lett.\  B {\bf 671}, 139 (2009)
  [arXiv:0807.3471 [gr-qc]].
\bibitem{Jackson:2009mz}
B.~M.~Jackson, A.~Taylor and A.~Berera,
  Phys.\ Rev.\  D {\bf 79}, 043526 (2009)
  [arXiv:0901.3272 [astro-ph.CO]].
\bibitem{Gavela:2009cy}
  M.~B.~Gavela, D.~Hernandez, L.~L.~Honorez, O.~Mena and S.~Rigolin,
  JCAP {\bf 0907}, 034 (2009)
  [Erratum-ibid.\  {\bf 1005}, E01 (2010)]
  [arXiv:0901.1611 [astro-ph]].
\bibitem{CalderaCabral:2009ja}
  G.~Caldera-Cabral, R.~Maartens and B.~M.~Schaefer,
  JCAP {\bf 0907}, 027 (2009)
  [arXiv:0905.0492 [astro-ph.CO]].
\bibitem{Valiviita:2009nu}
  J.~Valiviita, R.~Maartens and E.~Majerotto,
  Mon.\ Not.\ Roy.\ Astron.\ Soc.\  {\bf 402}, 2355 (2010)
  [arXiv:0907.4987 [astro-ph.CO]].
\bibitem{Majerotto:2009np}
  E.~Majerotto, J.~Valiviita and R.~Maartens,
  Mon.\ Not.\ Roy.\ Astron.\ Soc.\  {\bf 402}, 2344 (2010)
  [arXiv:0907.4981 [astro-ph.CO]].
\bibitem{Gavela:2010tm}
  M.~B.~Gavela, L.~Lopez Honorez, O.~Mena and S.~Rigolin,
  JCAP {\bf 1011}, 044 (2010)
  [arXiv:1005.0295 [astro-ph.CO]].
\bibitem{Honorez:2010rr}
  L.~L.~Honorez, B.~A.~Reid, O.~Mena, L.~Verde and R.~Jimenez,
  JCAP {\bf 1009}, 029 (2010)
  [arXiv:1006.0877 [astro-ph.CO]].
\bibitem{Martinelli:2010rt}
  M.~Martinelli, L.~Lopez Honorez, A.~Melchiorri and O.~Mena,
  Phys.\ Rev.\  D {\bf 81}, 103534 (2010)
  [arXiv:1004.2410 [astro-ph.CO]].
\bibitem{LopezHonorez:2010ij}
  L.~Lopez Honorez, O.~Mena and G.~Panotopoulos,
  Phys.\ Rev.\  D {\bf 82}, 123525 (2010)
  [arXiv:1009.5263 [astro-ph.CO]].
\bibitem{Refregier:2006vt}
  A.~Refregier {\it et al.},
  arXiv:astro-ph/0610062.
\bibitem{Refregier:2010ss}
  A.~Refregier, A.~Amara, T.~D.~Kitching, A.~Rassat, R.~Scaramella, J.~Weller and f.~t.~E.~Consortium,
  arXiv:1001.0061 [astro-ph.IM].
\bibitem{:2009pq}
    [LSST Science Collaborations and LSST Project Collaboration],
  arXiv:0912.0201 [astro-ph.IM].
\bibitem{Kodama:1985bj}
  H.~Kodama, M.~Sasaki,
  Prog.\ Theor.\ Phys.\ Suppl.\  {\bf 78}, 1-166 (1984).
\bibitem{Chongchitnan:2008ry}
  S.~Chongchitnan,
  Phys.\ Rev.\  {\bf D79}, 043522 (2009).
  [arXiv:0810.5411 [astro-ph]].
\bibitem{Corasaniti:2008kx}
  P.~S.~Corasaniti,
  Phys.\ Rev.\  {\bf D78}, 083538 (2008).
  [arXiv:0808.1646 [astro-ph]].
\bibitem{camb}
 A.~Lewis, A.~Challinor and A.~Lasenby,
  Astrophys.\ J.\  {\bf 538} (2000) 473.


\bibitem{Bartelmann:1999yn}
  M.~Bartelmann and P.~Schneider,
  Phys.\ Rept.\  {\bf 340} (2001) 291
  [arXiv:astro-ph/9912508].

\bibitem{Huterer:2010hw}
  D.~Huterer,
  Gen.\ Rel.\ Grav.\  {\bf 42} (2010) 2177
  [arXiv:1001.1758 [astro-ph.CO]].




\bibitem{Schneider:2002jd}
  P.~Schneider, L.~van Waerbeke, M.~Kilbinger and Y.~Mellier,
  Astron.\ Astrophys.\  {\bf 396} (2002) 1
  [arXiv:astro-ph/0206182].




\bibitem{Perotto:2006rj}
  L.~Perotto, J.~Lesgourgues, S.~Hannestad, H.~Tu and Y.~Y.~Y.~Wong,
  JCAP {\bf 0610} (2006) 013
  [arXiv:astro-ph/0606227].

\bibitem{Calabrese:2009tt}
  E.~Calabrese, A.~Cooray, M.~Martinelli, A.~Melchiorri, L.~Pagano, A.~Slosar and G.~F.~Smoot,
  Phys.\ Rev.\  D {\bf 80} (2009) 103516
  [arXiv:0908.1585 [astro-ph.CO]].


\bibitem{hirata:2003}
  C.~M.~Hirata and U.~Seljak,
  Phys.\ Rev.\  D {\bf 68} (2003) 083002
  [arXiv:astro-ph/0306354].


\bibitem{lensextr}
  T.~Okamoto and W.~Hu,
  Phys.\ Rev.\  D {\bf 67} (2003) 083002
  [arXiv:astro-ph/0301031].








\bibitem{Heavens:2003}
A. F. Heavens,
2003, MNRAS, {\bf 323}, 1327

\bibitem{Castro:2005}
P. G. Castro, A. F. Heavens, T. D. Kitching, 2005, Phys. Rev. D,
{\bf 72}, 3516

\bibitem{Heavens:2006}
A. F. Heavens,  T. D. Kitching, A. N. Taylor, 2006, MNRAS, {\bf
373}, 105

\bibitem{Kitching:2007}
T. D. Kitching, A. F. Heavens, A. N. Taylor, M. L. Brown, K.
Meisenheimer, C. Wolf, M. E. Gray, D. J. Bacon, 2007, MNRAS, {\bf
376}, 771


\bibitem{Hannestad:2006as}
  S.~Hannestad, H.~Tu and Y.~Y.~Y.~Wong,
  JCAP {\bf 0606} (2006) 025
  [arXiv:astro-ph/0603019].


\bibitem{art:Bacon} Bacon, D.; et al.; 2003; MNRAS, 363, 723-733

\bibitem{art:Massey} Massey R.; et al.; 2007, ApJS, 172, 239


\bibitem{art:Taylor} Taylor, A. N.; et al.; 2004, MNRAS, 353, 1176




\bibitem{Cooray:1999rv}
  A.~R.~Cooray,
  Astron.\ Astrophys.\  {\bf 348} (1999) 31
  [arXiv:astro-ph/9904246].
\bibitem{Smith:2002dz}
  R.~E.~Smith {\it et al.}  [The Virgo Consortium Collaboration],
  Mon.\ Not.\ Roy.\ Astron.\ Soc.\  {\bf 341} (2003) 1311
  [arXiv:astro-ph/0207664].

\bibitem{Abdalla:2007uc}
  F.~B.~Abdalla, A.~Amara, P.~Capak, E.~S.~Cypriano, O.~Lahav and J.~Rhodes,
  Mon.\ Not.\ Roy.\ Astron.\ Soc.\  {\bf 387} (2008) 969
  [arXiv:0705.1437 [astro-ph]].


\bibitem{Fu:2007qq}
  L.~Fu {\it et al.},
  Astron.\ Astrophys.\  {\bf 479} (2008) 9
  [arXiv:0712.0884 [astro-ph]].


\bibitem{:2006uk}
    [Planck Collaboration],
  arXiv:astro-ph/0604069.







\bibitem{Lewis:2002ah}
A. Lewis and S. Bridle,
Phys.\ Rev.\ D {\bf 66}, 103511 (2002) (Available from
\texttt{http://cosmologist.info}.)
\bibitem{Martinelli:2010wn}
  M.~Martinelli, E.~Calabrese, F.~De Bernardis, A.~Melchiorri, L.~Pagano and R.~Scaramella,
  Phys.\ Rev.\  D {\bf 83} (2011) 023012
  [arXiv:1010.5755 [astro-ph.CO]].
\bibitem{Tarrant:2011qe}
  E.~R.~M.~Tarrant, C.~van de Bruck, E.~J.~Copeland and A.~M.~Green,
  arXiv:1103.0694 [astro-ph.CO].





\end{thebibliography}
\end{document}